# The brain-AI convergence: Predictive and generative world models for general-purpose computation

Shogo Ohmae[1,*] & Keiko Ohmae[1]

[1] Chinese Institute for Brain Research (CIBR), Beijing, China

* Correspondence to: shogo@cibr.ac.cn

**Abstract**

**Recent advances in general-purpose AIs with attention-based transformers offer a potential window into how the neocortex and cerebellum, despite their relatively uniform circuit architectures, give rise to diverse functions and, ultimately, to human intelligence. This Perspective provides a cross-domain comparison between the brain and AI that goes beyond the traditional focus on visual processing, adopting the emerging perspective of world-model-based computation. Here, we identify shared computational mechanisms in the attention-based neocortex and the non-attentional cerebellum: both predict future world events from past inputs and construct internal world models through prediction-error learning. These predictive world models are repurposed for seemingly distinct functions—understanding in sensory processing and generation in motor processing—enabling the brain to achieve multi-domain capabilities and even human-like adaptive intelligence. Notably, attention-based AI has independently converged on a similar learning paradigm and world-model-based computation. We conclude that these shared mechanisms in both biological and artificial systems constitute a core computational foundation for realizing diverse functions including high-level intelligence, despite their uniform circuit structures. Our theoretical insights bridge neuroscience and AI, advancing our understanding of the computational essence of intelligence.**

# Introduction

AI development and brain research have a long history of mutual influence. In light of recent advancements in AI, there is growing expectation that studying the brain with reference to brain-like AI will lead to new theories and concepts in neuroscience [1-12]. In the past, each AI generally had a single function, but since 2018, general-purpose AIs have emerged that are capable of performing multiple functions with a single circuit. The cerebral neocortex and cerebellum, which together comprise more than 99% of all cells in the human brain, also achieve a wide range of sensory, motor, and cognitive functions through their relatively uniform circuit structures and local circuit computations [2,13-16]. Through comparison with AIs, we may be able to unravel the enduring mystery of how the versatile circuits in the neocortex and cerebellum realize a diverse array of functions. To address the mystery of the brain's universal local-circuit-level computations (hereafter, circuit computations), a comprehensive comparison of the brain and AI across various functional domains is essential. Such comparison is straightforward when there is a high degree of similarity in most aspects of the circuit computations; however, when there is only partial similarity, it is challenging to assess which aspects of the circuit computations are similar and to what extent, and to avoid merely listing fragmentary similarities. In fact, previous reviews comparing the brain and AI have been heavily biased toward the visual neocortex and brain-inspired vision AI because of their high degree of similarity, while failing to address representative neuroscience theories or historically successful AI circuits in other functional domains [1,3-7,11,17]. A comparison that focuses solely on visual processing is insufficient to satisfy the needs of AI experts seeking new insights from the brain, neuroscientists looking for novel perspectives from AI, or, more broadly, anyone curious about the secrets of human intelligence.

   To provide a comprehensive and structured comparison across domains, we break down the circuit-computation mechanisms into three elements and analyze the similarities between the brain and AI for each element. (i) Circuit architecture: In the brain, specific constraints on circuit architectures are imposed by the particular neuronal subtypes, the anatomical neuronal connections, and the learning sites at the connections. By contrast, the design of AI circuits is flexible and can be adjusted according to task demands. In both cases, the circuit structure is a critical factor that determines the upper limit of the computational capacity of the circuit [16,18]. (ii) Input/output signals: Information processing in the brain is fundamentally a transformation from input to output, so the input/output signals correspond one-to-one to the function of the circuit. (iii) Circuit learning: Even with similar input/output characteristics, different learning methods (e.g., supervised vs. unsupervised learning) can result in different intermediate processes across circuits [19-25]. By breaking down the circuit computation into these three elements and evaluating the similarities for each, we are able to comprehensively compare the brain and AI across sensory, cognitive, and motor domains for the first time. This new approach clarifies which aspects of circuit computation contain brain–AI similarities, allowing us to draw new insights from AI and relate these to key neuroscience theories and concepts.

In the context of macro-scale computation beyond local circuits, such as whole-brain processing, world models have recently gained considerable attention as a shared concept between the brain and AI. The world model (including modality-specific models, such as language models and visual world models) is an internal representation of the external environment, or a virtual simulator of the world, constructed by the brain and AI to understand and predict the external environment. A world model is not merely a collection of individual cases, but rather comprises generalized cases, including abstract features and structural rules, that support adaptation to novel situations [1,22,26]. The concept of world models has evolved in neuroscience and AI as follows. In neuroscience, the internal model theory proposed in the 1970s was established as a framework for explaining the predictive functions of the cerebellum in motor control, and has now become the standard theory for interpreting broad cerebellar functions (see Kawato, Ohmae, et al.) [16,27-40]. In the neocortex, the concept was introduced in the 2000s through research on predictive coding theory in sensory processing and model-based reinforcement learning [12,14,41-53]. In AI, large-scale language models and world-model-based reinforcement learning have gained increasing attention as core technologies in the pursuit of artificial general intelligence (AGI).

Considering the growing importance of the world model and its potential as a shared computational basis for both the brain and AI, we adopt the perspective of world-model-based processing to provide a comprehensive cross-domain comparison of their circuit computations. As a result, we identified a convergent evolution between the brain and modern AI, represented by large-scale language models. Based on this finding, we developed a new theory of macro-scale information processing mechanisms in the brain, corresponding to deep-layered local circuit computations: the neocortex and cerebellum acquire world models by predicting the future states of the world and learning from prediction errors, and use these predictive world models for a unified processing system across domains — understanding in sensory processing and generation in motor processing. This theory represents an integrated advancement of the internal model theory and the mirror neuron system, and indicates that world model-based circuit computation is key to enabling diverse functions within uniform circuit architectures in both AI and the brain. The approach and insights presented in this Perspective provide a promising foundation for future research bridging neuroscience and AI, with the aim of understanding human-like general intelligence.

# 1.1 Sensory information processing based on world models

## Sensory information processing in the neocortex

Research on sensory information processing—especially visual processing, which is crucial for humans and has been extensively studied—has led to influential theoretical proposals regarding input/output signals and learning algorithms in the neocortex. One representative theory is the unsupervised learning theory of the neocortex, proposed in 1999 (**Figure 1a**, green) [54]. This theory builds on the finding that processing in the primary visual neocortex can be explained by information compression performed by a sparse autoencoder trained via unsupervised learning, and the theory has been substantiated by experimental evidence (Supplementary Figure 1a,b) [4,17,23,54-57]. According to this theory, the input and output signals of the neocortex are essentially the same, with learning driven by minimizing reconstruction error (Supplementary Figure 1a, green). In addition, the hierarchical processing in the neocortex (e.g., in vision, the primary visual neocortex detects local edges and higher-order neocortices extract abstract features such as human faces) can be explained by a deep-layered sparse autoencoder (Supplementary Figure 1c) [58-61]. More recently, convolutional neural networks (CNNs), which are computationally more efficient than sparse autoencoders, have been used to approximate local circuits in the neocortex[1,3-7,11,17,56,57,62-68].

      Despite these successes, autoencoders and CNNs are feedforward circuit architectures and fail to capture the recurrent structure of the neocortical circuit, where diverse cell types exhibit complex recurrent connectivity that enables the integration of past and present information [69-72]. Consequently, although such neocortical theorization is sufficient for explaining static sensory processing (e.g., still image processing), it falls short in accounting for dynamic sensory processing in the neocortex (e.g., video processing) [16,70,73-83]. To address this limitation, the framework of neocortical unsupervised learning has evolved into the theory of predictive coding (Supplementary Figure 2) [46,84-86], which is now supported by abundant experimental evidence [8,85,87-93]. According to the predictive coding theory, the neocortex functions as a "predictor" of incoming sensory inputs from the external world, transforming these inputs into predictions and prediction errors (**Figure 1b**) [46]. Among these, the prediction error serves as a "newsworthy signal" and plays a central role in information processing. This prediction error is also used for learning in the predictor (i.e., the learning remains unsupervised). At the local circuit level, the neocortex has been approximated by RNNs such as long short-term memory (LSTM) networks, and their biological plausibility has been actively explored [94,95]. Within predictive coding circuits, intermediate layers perform information compression and feature extraction [69,71,85]. Moreover, the hierarchical structure of deep-layered predictive coding circuits

enables the progressive extraction and prediction of information at longer time scales and higher levels of abstraction (Supplementary Figure 2b-d) [12,14,85,86,96].

At the macro level, processing in the neocortical system serves two major functions: (1) to accurately predict future sensory inputs and acquire an internal model of the external world through unsupervised prediction-error learning; and (2) to enable sophisticated comprehension based on this model [12,14,42,43,48-52]. In essence, the neocortex acquires an internal model of the world via prediction-error learning, and this world-model-based processing underlies both high-level prediction and understanding.

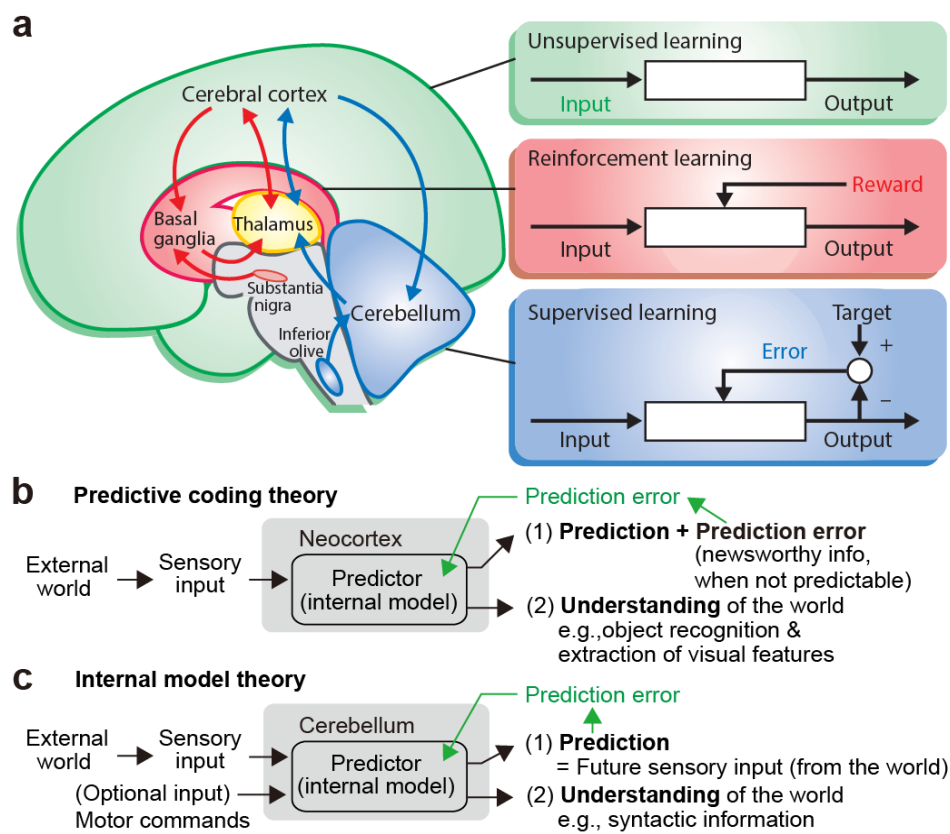

**Figure 1.** Theoretical frameworks of learning and processing in the neocortex and cerebellum. **a**. Unsupervised learning theory of the neocortex. Neocortical processing is established through unsupervised learning (green). In contrast, the cerebellum relies on supervised learning (blue), receiving high-level supervisory signals ("Target") from other brain regions. **b**. Predictive coding theory. The neocortex functions as a predictor, continuously receiving sensory inputs and (1) generating predictions of future inputs while outputting prediction errors as "newsworthy" signals. When the input matches the prediction, the output is zero, as the information carries no news value; only unpredicted inputs give rise to newsworthy signals. The prediction errors, in turn, serve as learning signals to update the predictor (green). Additionally, through information compression and feature extraction, this predictor contributes to (2) understanding of the world (e.g., visual feature extraction, object recognition, and 3D reconstruction). **c**. Internal model theory of the cerebellum. The cerebellum also functions as a predictor, receiving inputs from the external world, (1) generating predictions of future inputs, and learning from prediction errors. Accumulating evidence indicates that the learning target in the cerebellum is the future sensory input itself, rather than a high-level supervisory signal. Point (2) is detailed in **Figure 3b**. Panel **a** adopted from Doya (2000) with permission from Elsevier.

## Cerebellar sensory processing through internal world models

The cerebellum is densely interconnected with the neocortex in an area-by-area, parallel manner, and plays a supporting role in neocortical processing [15,97-99]. It is widely accepted that the cerebellum predicts future information using internal models of the world and that these internal models serve as the foundation of cerebellar processing (**Figure 1c**) [16,27-40]. According to this theory, the cerebellum acts as a predictor—or simulator—of the external environment, generating predictions of future inputs. In the context of sensory processing, the function of the cerebellum—its input-output transformation—is to predict future sensory signals from the external world on the basis of past sensory inputs. Learning in the cerebellum, as in the neocortex, is driven by prediction-error learning. Therefore, interestingly, the cerebellar input/output signals and learning can be regarded as equivalent to those of the neocortex. Traditionally, the cerebellum has been considered a supervised learning machine (**Figure 1a**, blue), in contrast to the neocortex as an unsupervised learning system [4,17,23,37,38,54-56]. However, the cerebellar teaching signals are simply prediction errors and do not necessarily contain highly processed information like correct labels of supervised learning in machine learning [37-40,97,100-102]. In this sense, we emphasize that cerebellar learning is essentially unsupervised prediction-error learning, as in the neocortex and machine learning. In contrast, the circuit architecture of the cerebellum differs markedly from that of the neocortex. Recurrent connections in the cerebellum are confined to specific cell types, resulting in a much simpler RNN circuit compared to the neocortex (below). The number of neurons in the cerebellum is almost four times greater than in the neocortex and cerebellar neurons are characterized by faster temporal responses. These features—its simpler yet large-scale architecture and high processing speed—are thought to underlie the cerebellar role in supporting neocortical processing [15,97-99].

# 1.2 Sensory language processing based on world models

Sensory language processing, i.e., language comprehension, consists of a stepwise process, in which speech is perceived, phonemes are recognized, words formed from these phonemes are identified, and the meanings of the words are integrated according to grammatical rules to derive the overall meaning of the sentence. Here, we focus in particular on language comprehension after word recognition, a high-level cognitive processing step characteristic of humans, to compare the brain and rapidly advanced language AI.

## Neocortical language processing through hierarchical predictive coding

Like visual processing, the neocortical language pathway is also hierarchically organized, with information flowing from lower-order language areas (e.g., STS) to higher-order language areas (e.g., IFG, angular, supramarginal). Language processing progresses in stages—from word recognition and semantic processing, to syntactic processing of word sequences, and ultimately to sentence comprehension [93,103-110]. With regard to learning and how the brain acquires language, the innate theory of language (in which basic language functions are innate and genetically determined) has historically been favored [111-113], but recent computational neuroscientific work has challenged this view, demonstrating that language functions that had been assumed impossible to learn can in fact be acquired through biological postnatal experiences and unsupervised learning [114,115].

     Research on human neocortical language processing has advanced dramatically since 2021, yielding two important insights. First, neocortical language processing can be explained by extending the hierarchical predictive coding framework of sensory processing to language inputs. Specifically, neocortical signals during language comprehension contain both prediction signals and prediction error signals for the next word [116,117], and these predictions are organized in a hierarchical structure where higher-order language areas make more abstract predictions, such as predicting the semantic category of the next word [93,118,119]. This extraction of increasingly abstract language information aligns naturally with the theoretical framework of hierarchical predictive coding in language [12,53,120]. Second, predictive coding theory predicts that neocortical signals should resemble those of hierarchical language AI trained on next-word prediction; indeed, neocortical signals most closely resemble such AI systems when compared against other language AI systems [107,116]. However, the match is not perfect: neocortical signals are better explained by the predictive coding framework, which predicts words beyond just the immediate next word [93].

Collectively, these findings indicate that neocortical language processing operates as hierarchical predictive coding, with lower layers performing simple word prediction and higher layers predicting more abstract content. Similar to sensory information processing, the neocortex acquires a world model of language through prediction-error learning, then uses this language model for both prediction and understanding by extracting abstract information. This represents the core computational principle shared between language processing in the neocortex and language models in AI, a parallel we explore in depth below.

## Large-Scale Language Models in AI: deep-layered RNN (GNMT) and deep-layered Transformers (BERT, GPT)

The first major milestone in recent advances in language AI was the development of Google Neural Machine Translation (GNMT), which was adopted by Google Translate in 2016 (**Figure 2a**) [121,122]. The GNMT circuit primarily consists of a deep-layered RNN incorporating LSTM. GNMT relies on supervised learning with human-made translation sentences, and generates errors only in the final layer, propagating them to all layers by backpropagation. While GNMT differs from the neocortex in these aspects (neocortical learning is unsupervised learning, and prediction errors in predictive coding are generated and consumed at each layer), it is noteworthy the GNMT encoder can convert a sentence in one language into a semantic vector (the population activity of a group of neurons), and this vector contained enough meaning of the original sentence for the decoder to generate a sentence with the same meaning in another language. This demonstrates that the deep-layered RNNs have the capacity to understand and convert sentences into semantic population-coding vectors at a level close to that of humans, which was groundbreaking from the neuroscience perspective [123,124].

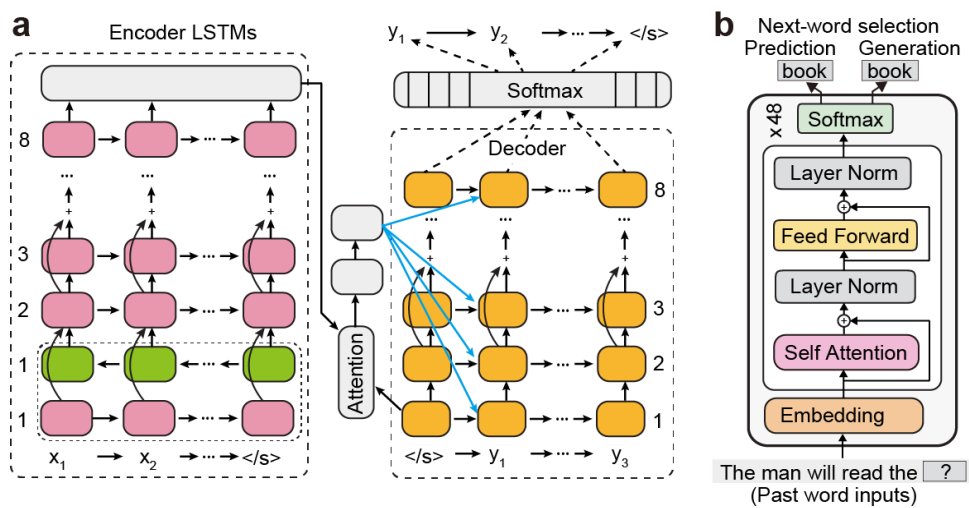

**Figure 2. Language-processing AI and analogy to the neocortex.**

**a,** The translation circuit of Google Neural Machine Translation (GNMT). This circuit consists of encoder, attention, and decoder units, with the encoder–decoder combination used only once throughout the entire circuit. Left, The encoder is a stacked-LSTM circuit that sequentially receives the words of a source sentence and integrates them to generate the signal that represents the meaning of the entire sentence. The encoder has eight layers (one bidirectional layer plus seven unidirectional layers), and the outputs of the final layer are combined to generate a vector representing the meaning of the entire sentence. The bidirectional layer ensures that the representation for even the first words can include the context of the sentence. Center: The attention unit is a two-layer neural network that receives the most recently translated word from the decoder (arrow from the decoder; only the signal flow during the second word [y2] translation is illustrated) and the vector representation of the source sentence from the encoder (arrow from the encoder), and then creates a weight-coefficient signal indicating which word in the source sentence should be focused on (i.e., the additive attention). The attention unit then provides this weight-coefficient multiplied source-sentence signal to each layer of the decoder (blue arrows to the decoder). The decoder is an eight-layer LSTM that sequentially generates the probabilities of each word being the next translated word. The circuit is trained through supervised learning with human-made translation examples as the correct targets. The error signal, calculated by comparing the generated word with the correct word, is delivered to the entire encoder–decoder system via backpropagation to update the synaptic weights. **b**, Transformer circuit for GPT-2. Since GPT is a circuit that predicts the next word, it does not have an encoder–decoder structure like GNMT but has an integrated structure. During training, the input is a sentence with one word masked. GPT makes a prediction for the next word (e.g., book). After training, for sentence generation, the input is an unfinished sentence created immediately before by GPT, and GPT repurposes the same output system to generate the next word. Circuit: Repeating blocks mainly consisting of an attention unit and a two-layer feedforward neural network, with these blocks stacked 12 times in GPT-1 and 48 times in GPT-2. Input/output: The input is a word sequence in a sparse representation (the maximum text length is 2048 words). The output is a probability representation of the next word candidate. Learning: Unsupervised learning through next-word prediction. Panel **a** adapted from Ramsundar & Zadeh (2018) with permission from O'Reilly Media. Panel **b** adopted from Wu Y. et.al (2024) (CC BY 4.0), with elements from Radford A. et al. (2018).

After GNMT, due to limitations in computational cost and speed (low efficiency of parallel processing with GPUs), AI language processing shifted from sequentially inputting words into an RNN to feeding entire sentences into a feedforward circuit known as a transformer [125,126]. Although this input format differs from that of the brain, the transformer expands the input dimensions to include the time axis, enabling the integration of information over time, similar to an RNN. In addition, interestingly, the prevailing learning method also transitioned from supervised learning to unsupervised learning, similar to learning in the brain.

The first to demonstrate the power and versatility of unsupervised learning in a transformer circuit was Google's BERT (bidirectional encoder representations from transformers)[20]. Pre-BERT language AIs were limited by the lack of versatility in supervised learning (i.e., accuracy dropped significantly outside of their trained tasks). To overcome this limitation, BERT, with an encoder–decoder circuit, employed large-scale unsupervised learning of masked-word completion. Subsequently, by replacing the decoder part with a small output layer for specific tasks, supervised learning of only the small output layer was sufficient to achieve state-of-the-art performance. The success of this method on a wide range of language tasks indicates that the encoder part of BERT is capable of generalized language understanding sufficient for various tasks.

Later, the power of unsupervised prediction-error learning was demonstrated sensationally by the GPT (generative pre-trained transformer) series, including ChatGPT (**Figure 2b**) [127-129]. Since GPT-2 in particular, GPTs have been thoroughly trained on next-word prediction. Remarkably, GPTs can repurpose their predictive ability of generating a future word to repeatedly generate the next word for sentence generation. This capability enables GPTs to generate coherent sentences, making them versatile enough to verbally answer a wide range of language tasks with top-level accuracy, without the need for the supervised fine-tuning required by BERT (e.g., in the task of scoring a product comment, when GPT is asked, "On a scale of 1 to 10, what is the positivity of this comment?", GPT can score it by generating a text sentence, like "This is an 8", whereas BERT needs a number-specific output unit). In terms of language comprehension, this demonstrates that a transformer trained with unsupervised prediction-error learning is capable of acquiring general-purpose language-understanding abilities (the text-generation ability is discussed in section 2.1).

All of these language AIs can appropriately interpret sentences with complex syntax encountered for the first time, indicating that they have acquired language models sufficient for language comprehension. Among them, GPT is particularly notable because its language model is acquired through prediction-error learning, which is thought to endow it with a language simulation capability analogous to that of the brain [107,115-117,119]. Indeed, GPT's signals more closely resemble neocortical signals than do those of BERT or GNMT [116].

# Processing similarities between the neocortex and the transformer circuits

Here, we compare the neocortex and transformer circuits, highlighting the high degree of similarity in their processing mechanisms (see Supplementary Figure 3 for circuit design and processing of the transformer). First, both the neocortex and transformers excel at integrating information over time. Second, both are circuits with relatively uniform structures that exhibit high versatility. Transformer circuits have excelled in various fields, achieving top accuracy in language processing and visual information processing (e.g., vision transformer), as well as excellent performance in video generation (e.g., DALL-E), suggesting their potential as a general-purpose circuit [22,125,130], similar to the versatile neocortical circuit [13,16]. Third, through large-scale prediction-error learning, powerful world models can be built by both the neocortex [12,14,41-53] and the transformer [126-128,131,132]. Fourth, they both possess attentional processing mechanisms. Transformer circuits were designed with reference to the attention mechanisms of the neocortex [18,133], particularly top–down attention to spotlight specific information depending on context. To summarize, both the neocortex and the transformer are versatile attentional circuits that excel at integrating information over time and have the ability to develop models of the external world through prediction-error learning.

The differences lie in the circuit structure. First, transformers are feedforward circuits that receive inputs all at once, whereas the neocortex operates as an RNN circuit, receiving inputs sequentially. However, this difference may not be fundamental, as any RNN can be represented by a time-unfolded feedforward circuit (e.g., **Figure 2a**). Indeed, RNNs with attention mechanisms equivalent to those of the transformer have been developed [6,134]. Second, transformers lack a unit corresponding to an artificial neuron. Yet, this may also not be a major difference, as the transformer's attention mechanism (dot-product attention) can be reformulated as a two-layer neural network implementing additive attention (Supplementary Figure 3). Furthermore, the integration of information by attention mechanisms in transformers resembles that of CNNs, which use artificial neurons, with the key difference being that whereas CNNs perform fixed-pattern integration of local inputs, transformers extend this to dynamic-pattern integration of distant inputs [135].

## Hierarchical attention-based processing as the core of neocortical processing

Traditional neuroscience implicitly assumed that attentional selection occurs only once during hierarchical processing in the neocortex, and research has sought to identify this "unique site." For example, there has been controversy about whether attention-based signal selection is performed at early or late stages of sensory processing [136,137] [138-140]. To resolve this debate, it has been proposed that the stage at which attentional selection occurs can be adjusted depending on task demands [141]. In addition, experiments have supported another theory—the network theory of attention—in which the oculomotor network, responsible for directing gaze to a particular direction, also plays an essential role in directing attention to that same direction [142-145]. However, to date, there is no definitive evidence supporting the assumption that attentional selection occurs only once in the neocortex, and the site of attentional selection has proved highly elusive. By contrast, in deep-layered transformer circuits, localized attention-based integration is performed at each layer. This raises an intriguing possibility: if the neocortex similarly performs attentional selection at each hierarchical level, it would naturally explain why the site of attention varies with task demands. Indeed, recent evidence supports this hypothesis. In human electrocorticography and fMRI studies, signals in higher-order language areas are best explained by the self-attention signals of deep-layered transformer circuits, directly suggesting that the neocortex implements hierarchical attentional processing analogous to that of deep-layered transformer circuits[107,146]. This indicates that the neocortex possesses an attention-based information-integration mechanism at each layer, making the neocortex a hierarchical attention-based processing circuit. Furthermore, given that deep-layered transformers specialized for attentional selection are sufficient to achieve or even surpass human-level performance in diverse tasks including language processing [22,125,130], hierarchical attention-based integration circuitry alone can be sufficient to serve as the core circuit computation underlying diverse neocortical functions. Although the attention function of the neocortex has traditionally been considered just one among many, our view suggests that it is the most foundational mechanism underlying neocortical processing.

## Cerebellar language processing: three-layer RNN with prediction-error learning

While the cerebellum is often considered a center for motor control, it is also responsible for a wide range of cognitive functions, including language processing [39,97,102,147-151]. The right lateral cerebellum (Crus I/II) is involved in two important non-motor language-processing functions: next-word prediction [152-155] and grammatical processing, particularly syntactic processing [151,156-159].

In cerebellar language processing, the three elements of circuit computation are considered as follows. Regarding the circuit architecture, although the cerebellum is often described as a typical feedforward circuit, there are abundant feedback projections from the output neurons (in the cerebellar nuclei; **Figure 3a**) to the input neurons (including both direct projections [160-163] and indirect projections [164-167]). These feedback projections are essential for the predictive functions of the cerebellum [163,168,169]. Regarding the input/output and learning, by extending the cerebellar internal model theory[16,28-35] to language processing, it has been proposed that the cerebellum receives sequential word inputs, generates a predictive output for the next word [152,153,170], and learns through the prediction-error signal for the next word [102,170].

Using these elements as the basis, we connected the three layers of input, Purkinje, and output neurons with feedforward and recurrent pathways to create a three-layered RNN circuit (**Figure 3a**). Interestingly, when this cerebellar artificial neural network was trained to perform next-word prediction, not only did the output of the circuit acquire the ability to predict the next word (**Figure 3b**, red arrow), but the intermediate layer of the word-prediction circuit (Purkinje cells) spontaneously acquired another cerebellar language function, syntactic processing (**Figure 3b**, blue) [171]. To the best of our knowledge, this is the first brain-imitating artificial circuit that aligns with all three elements of the circuit computation and that reproduces sophisticated human-characteristic cognitive functions.

Taken together, the cerebellum is approximated by a three-layer simple RNN circuit like Elman-type RNNs that has no attention mechanism and integrates information with a focus on the more immediate input. Similar to the neocortex and language AI processing, the cerebellar circuit acquires a language model via next-word prediction learning, and leverages it for both prediction and understanding by extracting abstract syntactic information (see **Figure 1c**).

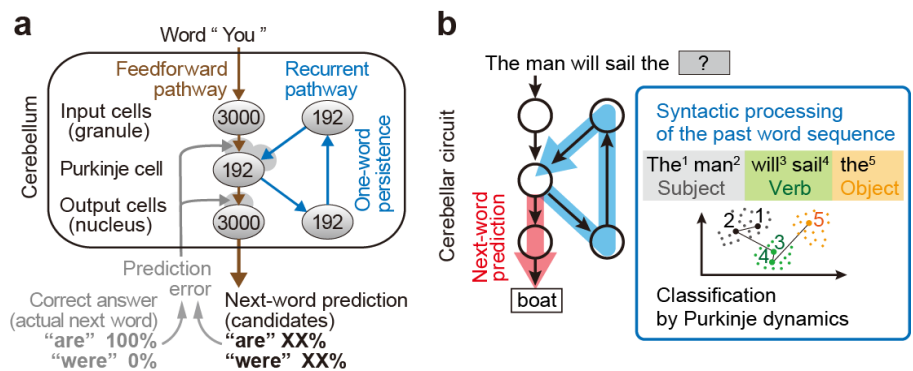

**Figure 3**. **Reproduction of cerebellar language functions by a three-layer artificial RNN imitating the cerebellum.**

**a**, A circuit was trained in a cerebellar language function—predicting the next word from a sequence of past words—through prediction-error learning. Each of the 3000 input neurons (granule cells) represents one word in the 3000-word set (sparse coding). Purkinje cells integrate the current word information in the feedforward pathway (brown) with the previous word-sequence information in the recurrent pathway (blue). The firing rates of the output neurons (cerebellar nucleus neurons) represent the probabilities that each of the 3000 words is the next word, forming the next-word prediction. The cerebellum receives the prediction-error signal from the inferior olive through the climbing fiber pathway (gray). This dedicated learning-signal pathway is a feature of the cerebellum, distinct from the neocortex. **b**, When the output neurons were successfully trained to predict the next word (red arrow), a circuit for syntactic processing (specifically, classification of words as subject, verb, or object) spontaneously emerged (blue arrow). Figure adapted from Ohmae K. and Ohmae S. (2024) (CC BY 4.0).

# 2.1 Motor language processing based on language models

Motor language processing, i.e., language generation, involves a step-by-step process of forming an idea to be conveyed, planning a sentence by selecting an appropriate combination of words and grammatical structures, and then outputting the resulting word sequence through vocalization or other actions. Here, we focus particularly on the sentence planning stage, as it represents a human-characteristic cognitive process.

## Integration of sensory and motor language processing in the brain and AI

Traditionally, the motor language-processing area (e.g., Broca's area) has been considered distinct from the sensory language-processing area (Wernicke's area). However, both sensory and motor language processing are now thought to be performed in significantly overlapping regions, including Broca's and Wernicke's areas in the neocortex and the right lateral cerebellum [104,108-110,152,153,157,172-174].

In language AI, GNMT and BERT have distinct sensory and motor processing units (corresponding to the encoder and decoder units), which aligns with the traditional understanding of language in the brain. On the other hand, in GPT, the circuits for sensory and motor processing significantly overlap. GPT also has an encoder–decoder structure, but these are integrated without a clear boundary. During training, GPT generates predictions (**Figure 2b**, top left), while establishing sentences comprehension. After training, in the sentence generation, this output system is directly repurposed to generate the next word (**Figure 2b**, top right). This overlap between sensory and motor processing in GPT aligns with more recent understanding of the brain.

This raises the question: does the brain similarly repurpose its word-prediction circuit for sentence generation? Recent studies of the human brain strongly support this hypothesis. In the neocortex, the network involved in word prediction during listening largely overlaps with that for sentence generation [110,116,175,176]. Interestingly, neuroscience has already identified a shared system for both listening and speaking—the mirror neuron system in language. In this system, the perceptual and predictive circuit of language is critically involved in language production (**Figure 4a**) [177-180]. Remarkably, this functional duality is confirmed even at the single-neuron level: human single-neuron recordings revealed that neurons involved in both listening and speaking exist in the prefrontal cortex [181]. In the cerebellum, the right lateral cerebellum has been demonstrated to perform the next-word prediction and grammatical processing, both of which can theoretically be viewed as language-model-based processing[171]. Since the right lateral cerebellum is also involved in planning for sentence generation [151,182], it is likely that the cerebellar word-prediction circuit is also used for sentence generation. Then, a striking commonality emerges: in AI, the neocortex, and the cerebellum, based on language models acquired through next-word prediction learning, three seemingly distinct functions—(1) word prediction, (2) language comprehension, and (3) language generation—are processed in a unified manner (**Figure 4b**).

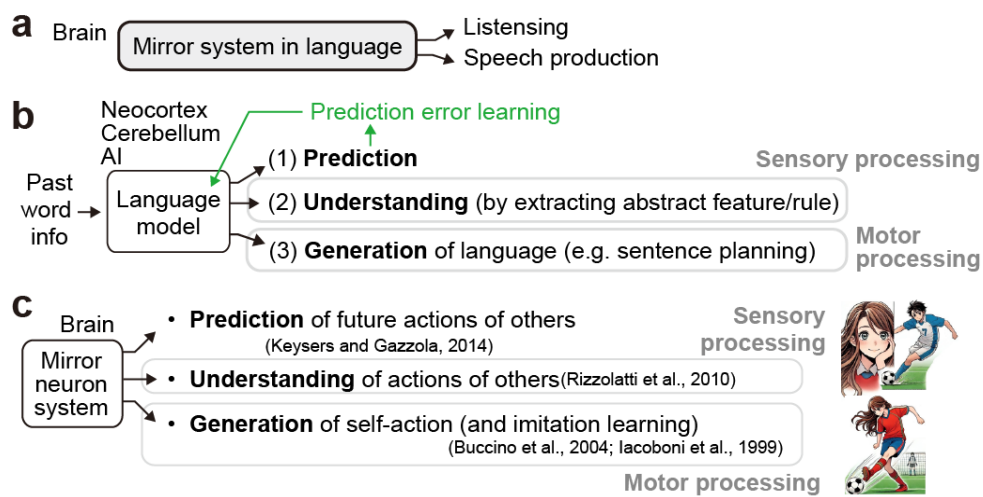

**Figure 4. World-model-based processing systems for prediction, understanding, and generation shared by the brain and AI.**

**a**, Mirror neuron theory in language processing. The Listening circuit, responsible for perceptual and predictive processing of language, is also crucially involved in language generation. **b**, Both the brain (neocortex and cerebellum) and AI acquire language models through next-word prediction learning, thereby establishing a unified processing system for prediction, understanding, and generation. **c**, The mirror neuron system in the brain can be viewed as a world model of actions. Acquired through prediction-error learning, it functions as a unified processing system that (1) predicts the actions of others, (2) observes and interpret actions of others, and (3) generates self-action planning.

## Reconceptualizing the mirror neuron system as a world model for actions

From the perspective of world models, the mirror neuron system in the motor domain—in which the same circuit for observing and predicting others' next actions is recruited for planning one's own actions —can be reconceptualized as utilizing a world model of actions (**Figure 4c**). The mirror neuron system is primarily located in the higher association areas of the neocortex and is acquired through the prediction of forthcoming actions of others [183]. Once established, this system engages in both sensory information processing (understanding the actions of others) and motor information processing (planning actions and imitation learning) [184-187]. This understanding of actions extends even to abstract representations of others' mental states underlying their actions, i.e., theory of mind [188]. Thus, the mirror neuron system utilizes a world model of actions—acquired through prediction-error learning—for prediction, understanding, and generation. In other words, the mirror neuron system in the neocortex can be regarded as an embodiment of an internal model of the world. On the other hand, the cerebellum has been understood within the framework of internal model theory but without assuming a mirror neuron system. If the cerebellum also realizes both sensory and motor processing within the same circuit, this suggests that a mirror neuron system exists in the cerebellum as well.

## Leveraging world models in generative AI for action planning

In AI, large language models have the ability to understand the actions and minds of others, and research to repurpose this ability to generate action plans (especially action planning for robots) has been rapidly flourishing since the release of ChatGPT at the end of 2022 [189]. For example, when a robot receives a human command and a goal in text format, the model provides a sequence of actions, in text format, suitable for achieving that goal. A representative example is Google's PaLM-SayCan [190]. When PaLM is asked to solve a problem, it describes the robot's situation and proposes strategies (i.e., sequences of actions) to resolve the problem. Subsequently, the controller, SayCan, evaluates the feasibility of the proposed first actions of these strategies, executes the most appropriate one, and updates its feasibility predictions according to the success or failure of the action. For such action planning, video-generative AI (e.g., a vision transformer, such as Stable Diffusion or DALL-E2) can provide candidate action sequences in video format (UniSim). Multiple methods employing video-generative AI for robot control have been proposed both for direct action planning (UniPi; Google, MIT; RFM-1; GR-1) and for imitation learning to train robot control (TRI). What these approaches in generative AI have in common is that they use predictive world models to understand the actions of others, and then repurpose the same system to generate their own actions. This structure closely parallels the "mirror neuron system as a world model" (above), indicating a profound correspondence—convergent evolution—between generative AI and the brain.

## Human-like adaptive language generation through in-context learning

Humans possess a remarkable ability: given appropriate instructions (in words and/or diagrams), they adapt immediately to novel tasks without additional training— a capacity rarely observed in other animals [191-193]. In the context of language generation, such tasks include translation, information provision, product-comment scoring, and so on. Recent language AIs (since GPT-3) have also achieved similar adaptive response capability, producing the famous commercially available AI, ChatGPT (released in 2022) [126,194,195]. Concretely, by receiving instructions called prompts, GPTs can adapt to a wide variety of new language tasks. This phenomenon, where the input–output transformation changes in response to the context of a given prompt, is called "in-context learning" [194,196,197]. Although the synaptic-weight parameters of the GPT circuit are updated only during training and remain unchanged when adapting to new tasks, the attention mechanism of the transformer retains the prompt-derived contextual signal and continuously propagates this signal throughout the circuit to change the input–output transformation — as if the parameters had been re-optimized in response to that prompt [196,198]. The emergence of in-context learning in GPT-3, achieved solely through next-word prediction, surprised even its developers. Recent analysis revealed that three factors are critical for the acquisition of in-context learning: prediction-error learning, the attentional computation of the transformer, and the large-scale size of the model [199-203].

While it has been suggested that there is a shared computational principle between GPT's in-context learning and human flexible adaptive capabilities [204], significant architectural differences exist between them. GPT lacks an explicit modular structure; consequently, there are no specific modules or layers that are particularly critical for in-context learning, and its acquisition is attributed to large-scale training of the large-scale attention circuit (Supplementary Figure 4) [126,194]. In contrast, in the human brain, the lateral prefrontal neocortex plays an essential role in this adaptive capability, and this region is proposed to serve as a hub that selects and switches between specialized neocortical "expert" regions (modules), each having mastered specific tasks, thereby enabling the quick handling of diverse tasks [96,193,205]. Intriguingly, recent evidence suggests that GPT also employs a form of expert specialization: distinct subcircuits function as hidden expert modules that are dynamically recruited for in-context learning by a multi-head self-attention mechanism [206,207]. Moreover, newer LLMs have explicitly incorporated the Mixture-of-Experts architecture—which, like the neocortex, selects and switches between specialized modules—into their transformer design (e.g., DeepSeek), and these LLMs achieve in-context learning performance comparable to GPT while using substantially smaller circuits [208,209]. Collectively, these findings suggest that the combination of attention mechanisms and Mixture-of-Experts constitutes a shared computational principle for in-context learning across both biological and artificial intelligence.

## Convergent evolution of language AI and human brain

The evolution of large language AI shows an intriguing convergence with human language-processing mechanisms. Initially, GNMT relied on supervised learning using human-translated texts. BERT then advanced to unsupervised learning (via masked-word completion), to acquire a general-purpose language-comprehension ability. GPT further pushed this evolution by switching to next-word prediction, endowing GPT with the sentence-generation ability that BERT lacked. This is because the next-word prediction mechanism could be transferred to next-word generation for sentence generation, whereas the BERT masked-word completion mechanism was difficult to apply to sentence generation. Interestingly, this evolution was not inspired by neuroscientific theories of language processing, but rather influenced them. The success of GPT has prompted a growing body of neuroscience studies indicating that next-word prediction contributes to language learning and processing in the neocortex [107,115-117,119]. Even more noteworthy is that this convergence is not limited to the learning algorithms, but extends to a macro-level processing system, in which predictive processing gives rise to a world-model-based processing system that enables entirely new capabilities such as understanding and generation (**Figure 4b**). Furthermore, combining prediction-error learning with attention mechanisms has enabled AI to acquire general adaptability to novel tasks, and the addition of the mixture-of-experts mechanism has made this acquisition more efficient. This likely reflects another point of convergence with the human systems that support similar intellectual flexibility. This multi-scale convergence reveals shared computational principles between brain and AI that underlie human-level intelligence. A deeper understanding of these principles will be central to bridging future neuroscience and AI.

# Discussion

## A new theory of a world-model-based processing system for diverse computations and intelligence

We have comprehensively compared the brain and AI across sensory, cognitive, and motor domains by subdividing circuit computations into three components—circuit architecture, input-output transformation, and learning algorithms—through the lens of world-model-based information processing. As a result, we found that recent AI circuits are converging toward brain-like processing, and obtained the following insights into the computational principles of the brain.

The neocortex and cerebellum predict future states of the external world from past inputs and learn to minimize prediction error. Through this process, they compress and abstract information about the external world to form compact, powerful world models. These models enable three fundamental types of information processing. (1) Prediction: Generating future information. (2) Understanding (sensory processing): Interpreting the external world by using abstract information within the world models, such as features, laws, and causal relationships (e.g., object recognition and text comprehension). (3) Generation (motor processing): Repurposing the future-information generation mechanism to generate other types of information output (e.g., action planning, language planning, and imitation learning). This world-model-based mechanism is the secret that allows the neocortex and cerebellum to accomplish a diverse range of functions despite their relatively uniform circuit structures and computations.

Moreover, while the input-output and, in particular, the learning algorithms of the neocortex and cerebellum have traditionally been considered conceptually distinct, we point out that they can, in fact, be regarded as fundamentally equivalent. Based on this view, we derive a common world-model-based processing system as described above. The differences between the two brain regions lie in their circuit computation abilities, as follows. The neocortex possesses complex and multiple recurrent connections and employs attention mechanisms to flexibly assign weights to temporal and spatial subsets of past inputs, resulting in dynamically changing patterns of integration that adapt to the context. In contrast to the traditional neuroscientific understanding of attention mechanisms, we propose that localized attention-based integration occurs at each hierarchical stage of neocortical processing, and that, by acquiring world models, these hierarchical attention circuits of the neocortex serve as the fundamental computational basis for powerful capabilities, including the immediate adaptability to novel tasks that characterizes human cognition. On the other hand, the cerebellum can be approximated by a large-scale yet simple three-layered RNN circuit, which integrates information with a fixed-pattern emphasis on the most recent inputs. It is widely believed that the neocortex and cerebellum cooperate to leverage their respective strengths in information processing [15,97,98]. Thus, we consider that the brain achieves sophisticated and rapid information processing through the cooperation of the neocortex, which performs complex and flexible computations, and the cerebellum, which specializes in simple but rapid processing.

In conclusion, world-model-based information processing is a common computational foundation that underlies a wide range of functions and intelligence in the neocortex, cerebellum, and AI. A deeper understanding of world-model-based computation will be essential not only for advancing and bridging neuroscience and AI, but also for illuminating the core principles of intelligence.

## Unsupervised prediction-error learning and world models in the brain

The idea that the neocortex and cerebellum acquire world models through unsupervised prediction-error learning provides an essential perspective for signal interpretation. For example, reward signals in these regions have been interpreted as learning signals for improving behavior in model-free reinforcement learning [37,39,210-218]. This is because animal behavior in psychological tasks—where animals can fully sample a limited set of task states—can be explained by classical model-free reinforcement learning that improves action selection based solely on the value of previously experienced actions and states [210-218]. However, the human brain also implements model-based reinforcement learning: it uses world models in the neocortex and cerebellum to simulate unexperienced situations [48,50,219,220] and determines action selection based on the inferred values of novel actions and states [48,221-224]. From the perspective of world models, a reward is simply one type of event to be predicted, alongside other world events. Reward-related signals can therefore be interpreted either as prediction signals of that event or as signals related to the prediction-error learning of that event (e.g., ground-truth signals and error signals). In other words, the reward signal in the brain is not necessarily limited to being the learning signal for improving behavior. In future research on the neocortex and cerebellum, the perspective of unsupervised learning of world models will lead to more careful and accurate interpretation of signals in the brain.

## Leveraging AI insights to understand neurological disorders

Insights from AI research can deepen our understanding not only of the brain but also of neurological disorders. One of the major mysteries of neurodevelopmental disorders is that their onset is probabilistic and not always reproducible. For instance, autism involves hundreds of genetic and non-genetic risk factors, with the interaction of multiple risk factors considered critical [225,226]. However, no decisive combination of interactions guarantees onset. By analogy with the fact that learning in AI circuits is also probabilistic, neurodevelopmental disorders may be viewed as probabilistic impairments in learning. In AI development, reducing the circuit size, improperly designing the circuit architecture, and choosing an unsuitable learning method destabilize the learning process and reduce the success rate. From this knowledge, we infer that in neurodevelopmental disorders, genetic and non-genetic risk factors destabilize learning and can result in the failure to achieve the appropriate developmental goal, ultimately leading to the onset of a disorder. By creating brain-imitating circuits and introducing manipulations to simulate these risk factors, we may significantly advance our understanding of the mechanisms underlying the onset of neurodevelopmental disorders.

## Implications for neuroscience of viewing the neocortex as a hierarchical attention circuit

Our perspective—that conceptualizing the neocortex as a hierarchical attention circuit is sufficient to explain its diverse and sophisticated functions—has the potential to reshape the framework of future research on neocortical computational mechanisms, which have traditionally been regarded as a collection of numerous distinct processes. For example, the function of the parietal neocortex is typically described in terms of two distinct roles: attention and movement intention [227,228]. However, recent advancements in AI transformer circuits, where attention and intention are integrated into a single circuit computation, suggest that these functions may represent different aspects of a unified process. Similarly, the prefrontal neocortex is currently associated with over ten distinct functions [229], but, considering that the attention mechanism in transformers can flexibly achieve apparently diverse functions through in-context learning, the fundamental processing in the prefrontal cortex can be better understood as a set of attention-based computations at high-level layers. Future neuroscience research is awaited to verify the possibility of reinterpreting the functions of various neocortical regions from the perspective of hierarchical attention circuitry.

## Scaling and the emergence of intelligence

The key factor behind recent advances in AI was the scaling up of simple circuits equipped with attention mechanisms alongside massive increases in training data [230-232]. The conventional belief was that overscaling of circuits results in too much capacity, leading to overfitting and reduced generalization to new data. However, in practice, large-scale AI circuits did not lead to overfitting. Instead, scaling up transformer circuits led to the emergence of in-context learning. A representative hypothesis for the underlying mechanism is the "lottery ticket" hypothesis (Supplementary Figure 4) [233,234], which proposes that (1) increasing the size of a circuit significantly raises the probability that it contains a subnetwork with complex processing capabilities prior to learning, and (2) after learning, only a small fraction of the circuit actually contributes to the output. Interestingly, corresponding observations have been reported in the human brain, with its vast number of synapses (~100 trillion). For point (1), there is already considerable similarity between the language signals in a pre-learning transformer circuit and those in the neocortex [116]. For point (2), weak synapses are pruned, and the associated molecular mechanisms have been identified (e.g., Arc's tagging of weak synapses) [235]. These parallels suggest that circuit scale expansion itself has contributed to the human brain's exceptional flexibility and intelligence relative to other species. The perspective that intelligence resides not in elegantly designed circuits but in large-scale attention circuits will be pivotal in advancing neuroscience. That said, the human brain learns far more rapidly than LLMs from vastly fewer examples and exhibits superior computational efficiency. As noted above, new-generation LLMs have successfully improved learning and computational efficiency by incorporating brain-like Mixture-of-Experts mechanisms, yet they remain far short of the brain's efficiency. In the process of bridging this efficiency gap, AI will continue learning from the brain, while neuroscience will gain new insights from AI.

## Beyond convergence: Illuminating intelligence through reciprocal development between neuroscience and AI

This Perspective integrated the scattered similarities between the brain and AI across multiple functional domains by examining their computational principles through the lens of world models—a concept common to both systems. We distilled their convergent evolution into three core principles: (1) prediction-error learning, (2) single-circuit computation of prediction, understanding, and generation, and (3) adaptability to novel tasks via attention mechanisms and Mixture of Experts. While our analysis centered on the neocortex and cerebellum, future research extending this comparative approach to other brain regions and functions—such as reinforcement learning in the brain and AI—promises to deepen our understanding of these functions and reveal new insights into the underlying computational mechanisms. Importantly, the identified convergence points serve as actionable guidelines for AI design: at these convergence points, aligning even more closely with the brain's computational strategies will enable the development of more efficient AI systems, while such brain-inspired AI will, in turn, substantially advance our understanding of biological intelligence—namely, the human brain. The deepening dialogue between neuroscience and AI promises to advance both disciplines, ultimately guiding us toward a fundamental understanding of intelligence.

**ACKNOWLEDGMENTS**

The authors thank lab members Chun Zhao and Tuo Xin for various contributions, including figure and reference editing and translation assistance, and Ouni Cao for his help with reference search. We also thank Mrs. Zoha Hassan for her meticulous editing, with attention to linguistic nuances. We extend our gratitude to Profs. Tatsuo Okubo (CIBR), Joji Tsunada (CIBR), and Takahiro Shinozaki (Science Tokyo) for their invaluable insights and comments that greatly enhanced the manuscript. Lastly, we thank Dr. Lindsay Bremner for her professional editing assistance. This work was supported by Chinese Institute for Brain Research (CIBR), Beijing.

**AUTHOR CONTRIBUTIONS**

S.O. conceptualized the study and prepared the initial manuscript and figures, integrating substantial input from K.O. Both authors collaboratively revised the text and figures and contributed to the finalization of the review manuscript.

**COMPETING INTERESTS**

The authors declare no competing financial interests.

# Supplementary Figures

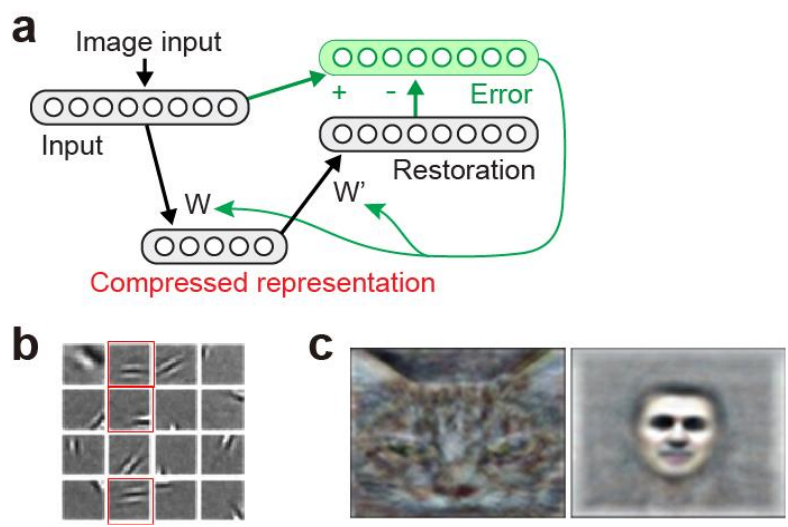

**Supplementary Figure 1. Deep-layered autoencoders as theories of visual information processing in the neocortex.** **a**, Autoencoder circuit. In a representative autoencoder circuit, the input neurons receive a 2D visual image (e.g., the pixel information of a photograph). The restoration neurons (downstream via disynaptic projections) are trained to encode the same information as the input neurons. If the downstream after the input neurons has fewer neurons, the input information needs to be compressed into a more compact dimensional space (red; compressed representation neurons). The error neurons calculate the difference between the input neurons and the restoration neurons, and the synaptic weights (W and W') of the circuit are updated to minimize this error. [Autoencoder] Circuit structure: This representative autoencoder circuit is a classical three-layer neural network (although various types of circuits can be trained to acquire the autoencoder function). Input/output: The input is a 2D visual image, and the output is the image reconstructed by the restoration neurons. Additionally, the signals from the compressed-representation neurons (red) serve as a secondary but more processed output (i.e., a de facto output). Learning: The synaptic weights (W and W') are updated to minimize the reconstruction error through unsupervised learning. The error signal (generated by the error neurons) is used solely for learning (green) and does not contribute to the outputs of the information processing. **b,** Neurons in the autoencoder circuit in **a** acquired receptive fields with local line/stripe patterns (described by Gabor functions, suitable for object contour/edge detection), as in the primary visual neocortex. The red frames highlight neurons with receptive fields that prefer horizontal lines/stripes in different local areas of the visual field. Furthermore, a phenomenon was observed in which the majority of receptive fields changed according to the characteristics of the training data, similar to the phenomenon in which vertical stripe receptive fields predominate in the visual neocortex of cats raised in a vertical stripe environment (Blakemore and Cooper, 1970). **c**, The optimal stimuli that most strongly activate two representative neurons in the highest layer of the deep-layered (hierarchical) autoencoder (Le et al., 2012). Left: The best activation stimulus for one neuron was this cat-like feature (Google's cat); this neuron is interpreted as a cat-classification neuron. Right: A human-face-classification neuron, whose best stimulus was a human facial feature, also emerged.
[Deep-layered autoencoder] Circuit structure: A deep circuit consisting of three repetitions of the receptive-field learning sub-layer (**a**) combined with a pooling sub-layer and a contrast-normalization sub-layer. Input/output: A deep-layered structure where the signal compressed in a lower layer is used as the input for the next layer. Learning: The three receptive field sub-layers are trained sequentially, one layer at a time, starting from the lowest layer, to reconstruct the input signal (greedy layer-wise training). Panel **c** adapted with permission from Le et al. (2012). Panel **b** adapted from Olshausen & Field (1996) with permission from Springer Nature. Panel **c** adapted with permission from Le et al. (2012).

# Deep-layered autoencoders as a circuit computation theory of the visual neocortex

To acquire models of the external world and understand the world, the brain generally needs to learn information processing with little instruction on the correct processing results [1,2]. The visual recognition circuit, located in the cerebral visual neocortex, is primarily trained through unsupervised learning [1-6]. In learning **theory** of the neocortex, unsupervised "autoencoder" learning is essential [2-4]. In autoencoder learning, the original input image serves as the target information to learn (**Supplementary Figure 1a**). In the autoencoder circuit, the input neurons receive a 2D visual image and the circuit is trained so that the restoration neurons restore exactly the same information as the input, i.e., the error between the two populations is minimized. It is called the autoencoder in machine learning terms, because it is trained to encode (ouput) the input itself. When the downstream neurons (red) is fewer then the input neurons, the input information needs to be compressed into a more compact dimensional space, resulting in the generation of compressed visual information as a secondary output.

The idea that the visual neocortex can function as an autoencoder has a long history [4]. In fact, when an autoencoder circuit of an artificial neural network (ANN) was trained with numerous natural images, the compressed-representation neurons acquired receptive fields similar to those of the primary visual neocortex (**Supplementary Figure 1b**) [7], suggesting that the primary visual neocortex also performs autoencoder-like feature extraction.

To acquire more high-level and abstract feature extraction, researchers created a deep-layered (hierarchical) autoencoder by stacking the data-compression functions of autoencoders, but training all layers simultaneously by backpropagation proved challenging. To overcome this challenge, in 2006, Hinton group proposed greedy layer-wise training, in which the deep autoencoder was trained sequentially from lower to upper layers [8,9], analogous to the critical period in the neocortices (early anchoring in lower visual areas) [10]. Using this learning procedure, in 2012, a group of researchers at Google successfully enabled neurons at the highest layer of a deep autoencoder to acquire specific responses to and recognition of human and cat faces (**Supplementary Figure 1c**). This was achieved by implementing tolerance sub-layers (more concretely, a pooling sub-layer and a contrast normalization sub-layer) in each layer, as first proposed in the "neocognitron" network [11], and by training the circuit through unsupervised autoencoder learning with large amounts of image data [12]. This result demonstrated the remarkable potential of the deep autoencoder to autonomously learn not only feature extraction of visual information but also object recognition and classification.

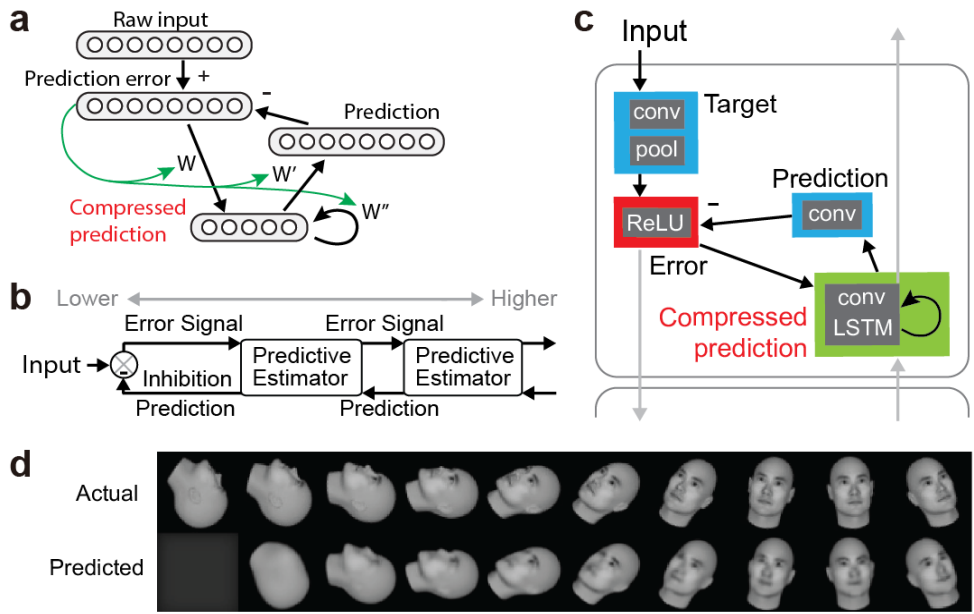

**Supplementary Figure 2. AI circuit similar to the predictive coding circuit, a processing theory for dynamic visual information in the neocortex.**

**a**, Structural diagram of a predictive coding circuit. Raw input information is processed by computing the difference between the input and the prediction, with only the resulting prediction error transmitted downstream. The compressed prediction neurons utilize a recurrent connection to integrate the current prediction-error signal (synaptic weight W) with the previous compressed prediction signal (W'') and transmit a new signal (W') to the prediction neurons. The prediction neurons restore the original data dimensions and predict the next input. The error signal (from the prediction-error neurons) is central not only for learning but also for information flow in the circuit. When the input is a static image, the past and new inputs are identical, and the predictive neurons reconstruct the input; this is equivalent to the autoencoder process. **b**, A schematic diagram of a deep-layered predictive coding circuit. Each layer (corresponding to the entire circuit in **a**) sends a copy of the error signal to the next higher layer and receives a copy of the compressed prediction from that higher layer. Since error signals are generated and consumed within each layer during learning, there is no need for backpropagation to transmit them to distant layers. Circuit: A raw input layer followed by a 3-layer recurrent neural network. Input/output: The input is a 2D visual image. The primary output is the prediction of the next input. With a deep-layered structure, secondary outputs are the prediction-error signal that is sent to the next higher layer and the compressed prediction sent to the lower layer. Learning: The synaptic weights of the circuit are updated to minimize the prediction-error signal (the learning signal is shown in green). **c,** Deep predictive coding network (PredNet). In terms of input/output signals and the learning process, PredNet faithfully implements the neocortical theory of deep predictive coding, incorporating the core circuitry shown in **a** with additional convolution sublayers, a pooling sublayer, and a long short-term memory (LSTM) unit. The LSTM integrates temporal information via a loop pathway (green). The inputs and outputs across layers are depicted with gray arrows. ReLU, rectified linear unit. **d,** Prediction by PredNet of the next frame in a sequence of 2D images (a face rotating in 3D space). PredNet successfully achieved higher prediction accuracy than either a CNN-RNN or an autoencoder tested by the authors.

Panel **b** adapted from Rao & Ballard (1999) with permission from Springer Nature. Panels **c,d** adapted with permission from Lotter W. et al. (2016).

**Evolution of circuit computation theory of the visual neocortex: Extension of autoencoders to time-directional information-integration circuits**

Deep-layered autoencoders and CNNs have succeeded in reproducing certain theoretical aspects of the circuit computations of the visual neocortex. However, they differ from the brain in their inability to integrate information in the time direction. Deep autoencoders and CNNs are feedforward circuits, and processing limitations of their structure prevent them from integrating past and current information, making them incapable of processing dynamic or sequential visual images (videos) [16,63-74]. By contrast, the neocortex, with multiple recurrent connections, is able to integrate past and present information and process dynamic information that changes over time [70,75-77]. To resolve this mismatch, autoencoder circuits need to be extended to be suitable for integration of information over time. The first candidate circuit structure is the RNN, which integrates information over time via recurrent pathways. (Transformer circuits, despite their feedforward structure, can also integrate information over time by extending the input format; discussed later.) For RNN input/output signals and learning, we propose "prediction-error learning"—where the circuit outputs predictions of future inputs on the basis of past inputs from the external world and learns from the prediction errors—for two reasons. First, predicting the future is equivalent to predicting dynamic changes in the external world; therefore, the circuit can acquire models of these dynamics through prediction-error learning. Second, prediction-error learning allows for numerous cycles of unsupervised learning, similar to the brain, which relies on extensive unsupervised learning [1,22,78-80].

    A theoretical circuit that fits these criteria has already been proposed for the neocortex: the predictive coding circuit (**Supplementary Figure 2**a) [31,81-83]. Predictive coding was originally proposed to explain receptive field properties that could not be explained by classical theories of the visual neocortex. A growing body of experimental evidence supports this theory [8,82,84]. The theory has also been generalized to other sensory areas, including the somatosensory and auditory neocortices [8,85-89]. The predictive coding circuit resembles the autoencoder circuit but differs in that the prediction-error signal is transmitted downstream and that the circuit with a recurrent pathway can receive inputs sequentially to integrate past and current information (**Supplementary Figure 2**a). When the input is a still image, the processing is equivalent to the autoencoder process, thus the predictive coding circuit can be interpreted as an RNN-oriented extension of the autoencoder, to enable information integration over time.

    Predictive coding circuits can be stacked to form a deep-layered structure to incorporate the hierarchical structure of the visual neocortices, with reciprocal connections between the layers [14,82,83,90]. In a deep-layered predictive coding circuit, prediction-error signals that cannot be predicted in a lower layer are sent to the next higher layer to become new prediction targets (**Supplementary Figure 2**b,c) [31]. This higher-level prediction signal is fed back as a top–down signal to the lower layer. Moreover, the free-energy principle, which mathematically formulates the prediction error in predictive coding circuits as entropy, focuses on the fact that the dynamics of

the world are hierarchical and proposes that the world can be modeled by mapping the dynamics onto a deep-layered predictive coding circuit [12,14].

**Prediction-error-learning RNNs as video processing AI: PredNet, CNN-RNN**

An AI circuit inspired by deep-layered predictive coding theory in the neocortex is the deep predictive coding network (PredNet) (**Supplementary Figure 2c**) [75,76]. PredNet is a practical implementation of the circuit theory of the visual neocortex, although its use of long short-term memory (LSTM) circuits for its RNN component does not replicate the circuit structure of the neocortex (LSTM is designed for efficient backpropagation and the effective integration of past and present information through gate functions). PredNet achieved high prediction accuracy for the next frame of rotating objects, such as a human face, in a 3D environment (**Supplementary Figure 2d**). Furthermore, it developed internal representations that efficiently coded object parameters (e.g., identity, view, rotation speed) in the intermediate layer. This demonstrates that PredNet replicates certain functional aspects of the neocortex that autoencoders could not.

PredNet exemplify RNN circuits, through prediction-error learning, can acquire world models that represent latent information absent in raw sensory data, such as 3D structures, by abstracting the information necessary for prediction. The neocortex also acquires internal models of the external world through experience [12,14,26-37,95] and extracts highly abstract sensory information at higher levels of the cortical hierarchy. This similarity indicates that the neocortex acquires world models through prediction-error learning, utilizing the compressed and abstract information within these models to interpret the external world.

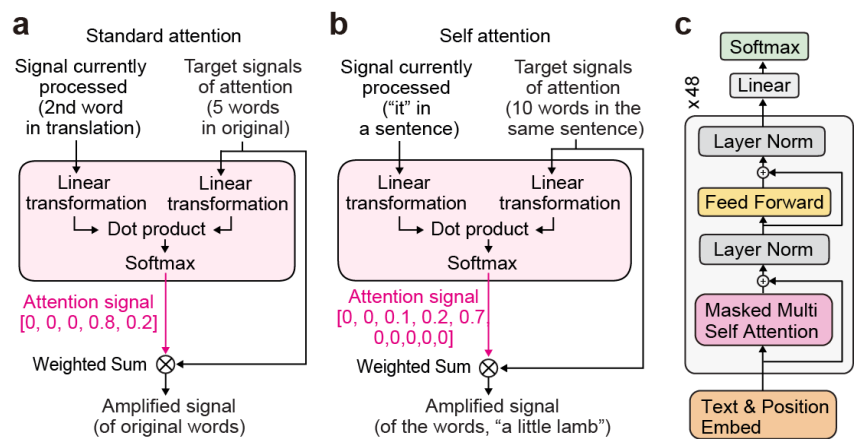

**Supplementary Figure 3.** Transformer attention circuit. **a**, Standard attention circuit. For simplicity, this will be explained with a concrete example. Consider the case where, during the translation of a sentence with five words, the signal for the second word that was most recently translated is used when generating the next word (input section). When selecting the next word, assuming that focusing on the corresponding word in the original sentence will lead to a more accurate and faithful translation, the attention mechanism (red box) generates an attention signal (degree of focus) for each word in the original sentence, with the sum being equal to 1 (red arrow; e.g., [0, 0, 0, 0.8, 0.2]). Next, this attention signal is used to weight the signal for each of the five words in the original sentence (with the attention signal serving as the weight coefficient), and the circuit outputs the original word information with the words of interest highlighted (output section). The first input is a high-dimensional signal (e.g., a 512-dimensional signal), which provides the context of the current processing. The second input is the five 512-dimensional signals corresponding to the five original words. In the attention mechanism of transformer predecessors (e.g., GNMT), this input/output was learned in a two-layer neural network (called additive attention; see **Figure 2a**). By contrast, the dot-product attention mechanism in the transformer assumes that this input/output is, in this case, converting the second translated word (the first input) into candidates for the third word, transforming dimensions from the original language to the target language, and then calculating importance by comparing it with each of the five original words (the second input). It also assumes that the conversion to the next word and the dimensional transformation can be done only by linear transformation (rotation and scaling operations in 512 dimensions), and the importance calculation is done via a dot-product operation (similarity calculation) (inside the red box). This dot-product attention mechanism, owing to the rather aggressive assumptions, is much faster in computation and learning than additive attention, yet, surprisingly, maintains comparable accuracy (to be precise, the linear transformation is applied to both input signals; in practice, to achieve higher performance, a multi-head attention process is employed, where eight linear transformations are performed in parallel to obtain eight parallel candidates of the attention signal). **b**, Self-attention mechanism. To understand "it" in a sentence, "Mary had a little lamb, and it was very white", the self-attention mechanism generates an attention signal that indicates which words in the sentence to attend to. The first and second inputs are signals from the same sentence (e.g., the signal for the word, "it", and the signal for the ten words of the sentence). The processing steps are the same as in **a**. If the word "it" corresponds to "a little lamb," the self-attention mechanism focuses on these three words and creates a signal to represent "a little lamb" by multiplying the attention coefficients (red arrow) and the word signals of the sentence (i.e., the copy of the second input) so that the output represents the detailed meaning of "it" (the information regarding "it" itself is also conveyed downstream by bypassing the attention mechanism; see the bypass path in **c**). **c**, Transformer circuit of GPT-2. Restatement of **Figure 2b**.

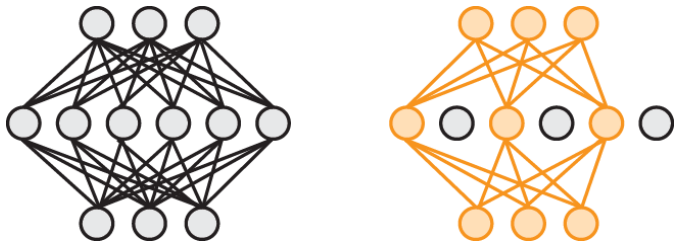

**Supplementary Figure 4.** Significance of scaling up the circuit size to intelligence and the "Lottery Ticket Hypothesis".

A representative hypothesis, known as the "lottery ticket hypothesis," proposes the following mechanism for how circuit scaling contributes to the acquisition of intelligence. First, an artificial neural circuit contains numerous subnetworks, and as the circuit is scaled up, the number of the subnetworks increases exponentially. For example, in a three-layer 3-6-3 network (black), the number of 3-3-3 subnetworks (orange) is Combination(6, 3) = 20. These subnetworks overlap, but no pair has exactly the same initial value combination). During learning, a subnetwork whose initial values happen to be well-suited for task-demanding processing contributes to the output and is reinforced, eventually playing a major role in processing of the entire circuit. Therefore, larger circuits are more likely to contain such a "hit" subnetwork and to be successfully trained. This hypothesis is experimentally supported [31,32].

The lottery ticket hypothesis implies two points: (1) As the circuit scales, the number of subnetworks increases sharply, boosting the likelihood of containing subnetworks with complex processing capabilities. (2) Even if a large circuit is being trained, only a small fraction of the circuit actually contributes to the output generation after learning (This is called "implicit regularization" because it is as if a rule that facilitates learning in only a small fraction of the synapses has been added). Point (1) suggests that when learning sophisticated information processing like in-context learning, which requires complex processing mechanisms (see "Immediate Adaptability to Novel Tasks" in motor language system)[33,34], the existence of a pre-learning subnetwork that happens to be well-suited for such complex processing beforehand is essential for successful learning. In line with (2), it was found that only a small portion of the vast number of synapses in the large AI circuit are used after training, and that even if 90% or more of weak synapses are pruned (i.e., further reduced to zero), most of the performance gained from learning can be maintained [35-38]. This in turn means that a large-scale circuit is necessary only during the learning phase [31,32].

**The importance of circuit scale in neuroscience**

The fact that large-scale circuitry in AI was a key factor in acquiring advanced processing capabilities had a profound impact on AI research and computational neuroscience, where researchers had been focusing on designing intelligent circuits utilizing sophisticated information theory and competing to improve their performance (Sutton, "The Bitter Success")[39]. This finding has two implications for future neuroscience research.

First, with the relative ease of creating artificial circuits imitating the brain, neuroscience research combining experiments and artificial neural networks will become increasingly prominent. Conventional circuits designed by computational neuroscientists were based on sophisticated theories and were often esoteric to other neuroscientists, but the advent of CNNs and transformers has made it possible to compare the brain with simpler (albeit large-scale) artificial neural circuits. As a result, the use of artificial neural networks is becoming an essential approach across neuroscience research. That said, the information processing in AI circuits remains a black box, and it is largely unclear what specific steps and intermediate stages are taken to process information. As we move forward, the rich reservoir of neuroscience theories on computational mechanisms with sophisticated designs will play a crucial role in understanding the internal processing of AI circuits and brain-like circuits.

Second, the emergence of immediate adaptability to novel tasks through the scaling up of circuits and training data in AI will raise awareness of the importance of circuit scale in neuroscience research. For example, this AI finding suggests that to elucidate sophisticated human-characteristic processing, human brain circuits, which are far larger than those of mice or monkeys, need to be examined. On the other hand, to elucidate the design principles of brain circuits, it is beneficial to study non-human brains, which are similar in design to the human brain, albeit on a smaller scale. Although the view that intelligence lies not in cleverly designed circuits but in large circuits has not been given much weight in neuroscience, awareness of the scale of circuits will be essential for future research on human intelligence.